\newif\ifsingle

\newif\ifFullVersion
\FullVersiontrue 

\ifsingle
\documentclass[12pt,draftclsnofoot, onecolumn]{IEEEtran}		
\else		
\documentclass[10pt,final, twocolumn]{IEEEtran}
\fi

\usepackage{times}
\usepackage{amsmath,dsfont}
\usepackage{amssymb,amsthm}
\usepackage{epsfig,verbatim}
\usepackage{subcaption}
\usepackage{setspace}
\usepackage{color}
\usepackage{cite}
\usepackage{epstopdf}
\usepackage{graphics}
\usepackage{accents}
\usepackage{acronym}
\usepackage{booktabs}
\usepackage{mathtools} 
\usepackage{enumitem}
\usepackage{multirow}
\usepackage{dblfloatfix}
\usepackage[bookmarks,colorlinks]{hyperref}
\usepackage{gensymb}
\usepackage[flushleft]{threeparttable}

\usepackage[ruled,linesnumbered]{algorithm2e}  

\SetKwInput{KwData}{\textbf{Init}} 
\let\oldnl\nl
\newcommand{\nonl}{\renewcommand{\nl}{\let\nl\oldnl}}

\ifsingle

\newcommand{\figSpace}{\vspace{-0.2cm}}

\else

\newcommand{\figSpace}{\vspace{-0.2cm}}
\fi 

\acrodef{adc}[ADC]{analog-to-digital convertor}
\acrodef{cs}[CS]{compressed sensing}
\acrodef{dtft}[DTFT]{discrete-time Fourier transform}
\acrodef{dnn}[DNN]{deep neural network} 
\acrodef{mc}[MC]{monte carlo}
\acrodef{csi}[CSI]{channel state information}
\acrodef{bpsk}[BPSK]{binary phase shift keying}
\acrodef{qpsk}[QPSK]{quadrature phase shift keying}
\acrodef{map}[MAP]{maximum a-posteriori probability}
\acrodef{snr}[SNR]{signal-to-noise ratio}
\acrodef{bs}[BS]{base station} 
\acrodef{iot}[IOT]{Interent of Things}
\acrodef{mmtc}[mMTC]{massive machine-type communications}
\acrodef{embb}[eMBB]{enhanced Mobile Broadband}
\acrodef{mimo}[MIMO]{multiple-input multiple-output}
\acrodef{siso}[SISO]{single-input single-output}
\acrodef{mse}[MSE]{mean-squared error}
\acrodef{pdf}[PDF]{probability density function}
\acrodef{rv}[RV]{random variable}
\acrodef{ml}[ML]{machine learning}
\acrodef{fec}[FEC]{forward error correction}
\acrodef{rs}[RS]{Reed-Solomon}
\acrodef{ar}[AR]{augmented reality}
\acrodef{vr}[VR]{virtual reality}
\acrodef{lti}[LTI]{linear time-invariant}
\acrodef{wss}[WSS]{wide-sense stationary}
\acrodef{psd}[PSD]{power spectral density}
\acrodef{ser}[SER]{symbol error rate} 
\acrodef{ber}[BER]{bit error rate} 
\acrodef{gd}[GD]{gradient descent}
\acrodef{sgd}[SGD]{stochastic gradient descent} 
\acrodef{isi}[ISI]{intersymbol interference}  
\acrodef{awgn}[AWGN]{additive zero-mean white real Gaussian noise} 
\acrodef{ut}[UT]{user terminal} 
\acrodef{mmw}[mmWave]{millimeter wave}
\acrodef{noma}[NOMA]{non-orthogonal multiple access}
\acrodef{mac}[MAC]{mulitple access channel}
\acrodef{fl}[FL]{Federated learning}
\acrodef{lstm}[LSTM]{long short-term memory}
\acrodef{maml}[MAML]{model-agnostic meta-learning}
\acrodef{sic}[SIC]{soft interference cancellation}
\acrodef{pmf}[PMF]{probability mass function}
\acrodef{urllc}[URLLC]{ultra-reliable and low-latency communication}
\acrodef{sova}[SOVA]{soft-output Viterbi algorithm}
\acrodef{wbp}[WBP]{weighted belief propagation}
\acrodef{ecc}[ECC]{error-correction codes}
\acrodef{crc}[CRC]{cyclic redundancy check}
\acrodef{scl}[SCL]{successive list cancellation}
\acrodef{bp}[BP]{belief propagation}
\acrodef{mle}[ML]{maximum likelihood}
\acrodef{sota}[SOTA]{state of the art}
\acrodef{irs}[IRS]{intelligent reconfigurable surface}
\acrodef{OAMP}[OAMP]{orthogonal approximate message passing}
\acrodef{bcjr}[BCJR]{Bahl-Cocke-Jelinek-Raviv}
\acrodef{ai}[AI]{artificial intelligence}
\acrodef{nlp}[NLP]{natural language processing}
\acrodef{rnn}[RNN]{recurrent neural network}

\IEEEoverridecommandlockouts 

\title{Adaptive and Flexible Model-Based AI for Deep Receivers in Dynamic Channels}
\author{  
	\IEEEauthorblockN{Tomer Raviv, Sangwoo Park,  Osvaldo Simeone, Yonina C. Eldar, and Nir Shlezinger
	} 
	\thanks{
		This project has received funding from the Israeli 5G-WIN consortium, the European Union’s Horizon 2020 research and innovation program under grants No. 646804-ERC-COG-BNYQ, as well as  725731, and by the European Union’s Horizon Europe project CENTRIC (101096379).  
        Support is also acknowledged for the Israel Science Foundation under grant No. 0100101, and for an Open Fellowship of the EPSRC with reference EP/W024101/1.
		T. Raviv and N. Shlezinger are with the School of ECE, Ben-Gurion University of the Negev, Beer-Sheva, Israel (e-mail: tomerraviv95@gmail.com; nirshl@bgu.ac.il).  
		S. Park and O. Simeone are with the Department of Engineering, King’s College London,  U.K. (email: \{sangwoo.park; osvaldo.simeone\}@kcl.ac.uk).
		Y. C. Eldar is with the Faculty of Math and CS, Weizmann Institute of Science, Rehovot, Israel (e-mail: yonina.eldar@weizmann.ac.il).}

}

\begin{document}

\maketitle

	\pagestyle{plain}
	\thispagestyle{plain}
	\begin{abstract} 
\Ac{ai} is envisioned to play a key role in future wireless technologies, with \acp{dnn} enabling digital receivers to learn  to operate in challenging communication scenarios.
However, wireless receiver design  poses unique challenges that fundamentally differ from those encountered in traditional deep learning domains. 
The main challenges arise from the limited power and computational resources of wireless devices, as well as from the dynamic nature of wireless communications, which causes continual changes to the data distribution. 
These challenges  impair conventional \ac{ai} based on highly-parameterized \acp{dnn}, motivating the development of adaptive, flexible, and light-weight \ac{ai} for wireless communications, which is the focus of this article. Here, we propose that AI-based design of wireless receivers  requires rethinking of the three main pillars of \ac{ai}: {\em architecture}, {\em data}, and {\em training algorithms}. In terms of  architecture, we review how to design compact \acp{dnn} via model-based deep learning. Then, we discuss how to acquire training data for deep receivers without compromising spectral efficiency. Finally, we review efficient, reliable, and robust training algorithms via meta-learning and generalized Bayesian learning. 
Numerical results are presented to demonstrate the complementary effectiveness of each of the surveyed methods. We conclude by presenting opportunities for future research on the development of practical deep receivers.
\end{abstract}
    \acresetall
	\section{Introduction}

Wireless communication technologies are subject to escalating demands for connectivity and throughput, with rapid growth in media transmissions, including images, videos, and,  in the near future, augmented and virtual reality. Furthermore, transformative applications such as the \ac{iot}, autonomous driving, and smart manufacturing are expected to play major roles in the new 5G-defined deployments of \ac{urllc} and \ac{mmtc} services. To accommodate these scenarios, communication systems  must meet strict performance requirements in  reliability, latency, and complexity~\cite{samsung202065}.

To facilitate meeting these  performance requirements, 
emerging technologies such as mmWave and THz communication, holographic \ac{mimo}, spectrum sharing, and \acp{irs} are currently being investigated. While these technologies may support  desired performance levels, they also introduce substantial design and operating complexity~\cite{samsung202065}. 
For instance, holographic \ac{mimo} hardware is likely to introduce non-linearities on transmission and reception; the presence of \acp{irs}  complicates channel estimation; and classical communication models may no longer apply in novel settings such as  the mmWave and THz spectrum, due to violations of far-field assumptions and lossy propagation.   This paper addresses the latter source of complexity by focusing on efficient design of receiver processing. 



Traditional receiver processing design is {\em model-based}, relying on simplified channel models, which, as mentioned, may no longer be adequate to meet the requirements of next-generation wireless systems. The rise of deep learning as an enabler technology for \ac{ai} has revolutionized various disciplines, ranging from computer vision and \ac{nlp} to speech refinement and biomedical signal processing. 
The ability of \acp{dnn} to learn mappings from data has spurred growing interest in their usage for receiver design in digital communications~\cite{dai2020deep,tong2022nine}. \ac{dnn}-aided receivers, referred to henceforth as \textit{deep receivers}, have the ability to succeed where classical algorithms may fail. Specifically, deep receivers can learn a detection function in scenarios having no well-established physics-based mathematical model, a situation known as {\em model-deficit}; or in settings for which the model is too complex to give rise to tractable and efficient model-based algorithms, a situation known as {\em algorithm-deficit}. 
Consequently, deep receivers have the potential to meet the constantly growing requirements of wireless systems. 

\setlength{\arrayrulewidth}{0.5mm}
\setlength{\tabcolsep}{22pt}
\renewcommand{\arraystretch}{1.4} 
\begin{figure*}[t]
\begin{minipage}{0.3\linewidth}
\centering
\includegraphics[width=\linewidth]{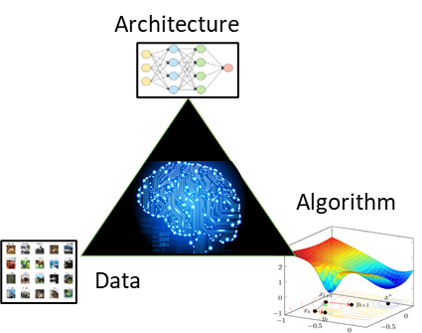}  
\end{minipage}
\hfill
\begin{minipage}{0.66\linewidth}
    \centering
      \begin{threeparttable}
      \small{
            \begin{tabular}{| c | c | c |}
                \hline
                \textbf{Pillar} & \textbf{Method} & \textbf{Literature}\\ 
                \hline
                \multirow{2}{4em}{Architecture} & Deep unfolding & \cite{he2018model,khani2020adaptive,shlezinger2019deepSIC } \\  
                 & DNN-aided inference & \cite{shlezinger2019viterbinet} \\
                 \hline
                 \multirow{2}{4em}{Data} & Self-supervised training & \cite{shlezinger2019viterbinet,fischer2022adaptive,schibisch2018online,finish2022symbol}\\
                 & Data augmentation & \cite{huang2019data,raviv2022data} \\
                 \hline
                 \multirow{3}{4em}{Training Algorithm} & Meta-learning & \cite{raviv2023online, goutay2020deep, chen2023learning,liu2022learning}\\  
                     & Bayesian learning & \cite{zecchin2022robust,raviv2023modular}\\
                     & Modular training & \cite{raviv2023online}\\
                \hline
            \end{tabular}
            }
      \end{threeparttable}
      \end{minipage}
              \caption{A summary of methods surveyed in this article that adapt the three pillars of AI to the requirements of deep wireless receivers.}
          \label{tab:methods}
\end{figure*}

Despite their promise, several core challenges  arise from the fundamental differences between wireless communications and traditional \ac{ai} domains such as computer vision and \ac{nlp}, 
 limiting the  widespread applicability of deep learning in wireless communications. The first challenge is attributed to the nature of the {\em devices} employed in communication systems. Wireless communication receivers are  highly constrained in terms of their computational ability, battery consumption, and memory resources. However, deep learning inherently relies on highly parameterized architectures, assuming the availability of powerful devices, e.g., high-performance computing servers.  

A second challenge stems from the nature of the wireless communication {\em domain}. Communication channels are dynamic, implying that the receiver task, dictated by the data distribution,  changes over  time. This makes the standard pipeline of data collection, annotation, and training highly challenging. Specifically, \acp{dnn} rely on  (typically labelled) data sets to learn from the underlying unknown, but stationary, data distributions.  For example, machine translation tasks, requiring the mapping of an origin language into  a destination language, do not change over time, enabling the collection of a large volume of  training data and the deployment of a pre-trained, static, \ac{dnn}. This is not the case for wireless receivers, whose processing task depends on the time-varying channel, restricting the size of the training data set representing the task.

The two challenges outlined above imply that successfully applying \ac{ai} for wireless receiver design requires deviating from conventional deep learning approaches. To this end, there is a need to develop communication-oriented \ac{ai} techniques, which are the focus of this article. 
Previous tutorials on \ac{ai} for communications, e.g., \cite{dai2020deep, tong2022nine}, have primarily concentrated on surveying diverse challenges and applications of conventional deep learning methods in the context of communication. In contrast, the present article aims to review approaches that address the unique challenges in the design of deep receivers that arise from the mentioned limitations of wireless devices and from the dynamic nature of the communication domain. Our main objective is  to provide a systematic review of  research directions  that target the practical deployment of deep receivers.

We commence by  motivating the development of \ac{ai} systems that are {\em light-weight}, and thus operable on power and hardware limited devices, as well as {\em adaptive and flexible}, enabling online on-device adaptation. As illustrated in Fig.~\ref{tab:methods}, we then propose that AI-based  wireless receiver design requires revisiting the three main pillars of \ac{ai}, namely  (\emph{i}) the {\em architecture} of AI models; (\emph{ii}) the {\em data} 
 used to train AI models; and (\emph{iii}) the {\em training algorithm} that optimizes the AI model for generalization, i.e., to maximize performance  \emph{outside} the training set (either on the same distribution or for a completely new one).

For each of these \ac{ai} pillars, we survey candidate approaches for facilitating the operation of the deep receivers. (\emph{i}) We first discuss how to design light-weight trainable architectures  via \emph{model-based deep learning}~\cite{shlezinger2022model}. 
This methodology hinges on the principled incorporation of model-based processing, obtained from  domain knowledge on optimized communication algorithms, within  AI architectures. (\emph{ii}) Next, we investigate how labelled data  can be obtained {without} impairing spectral efficiency, i.e., without increasing the pilot overhead. To this end, we show how receivers can generate labelled data by \emph{self-supervision} aided by existing communication algorithms; and how they can further enrich data sets via \emph{data augmentation} techniques that utilize 
 invariance properties of communication systems. (\emph{iii}) Finally, we cover training algorithms for deep receivers that are 
 designed to meet requirements in terms of efficiency, reliability, and robust adaptation of wireless communication systems, avoiding overfitting from limited training data while limiting training time. These methods include communication-specific \emph{meta-learning} as well as \emph{generalized Bayesian learning} and \emph{modular learning}.


To illustrate the individual and complementary gains of the reviewed approaches, we provide a numerical study considering finite-memory \ac{siso} channels as well as multi-user \ac{mimo} systems.  We conclude by discussing the road ahead, as well as  key research challenges that are yet to be addressed to enable  adaptive and flexible  light-weight deep receivers.

	\section{Deep Receivers in Dynamic Channels}
\label{sec:deep_receivers}

\begin{figure*}[t]
    \centering
    \includegraphics[width=1.6\columnwidth]{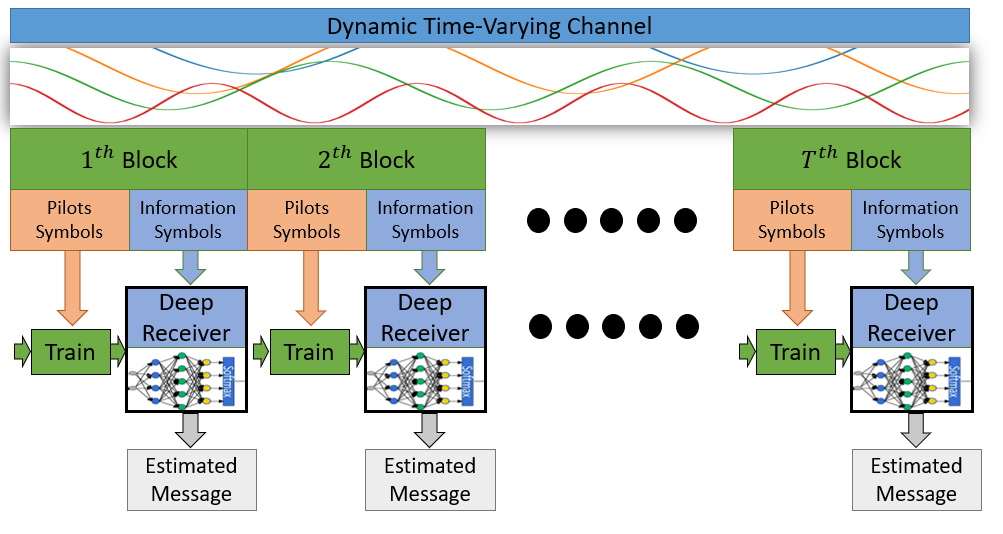}
    \caption{Overall illustration of online training of deep receivers in time-varying channels.}
    \label{fig:online_training}
\end{figure*}

As discussed in the previous section, harnessing the potential of deep learning in wireless systems requires communication-specific \ac{ai} schemes that are adaptive, flexible, and light-weight. The light-weight requirement follows from the power and computational constraints of wireless devices; while the need for adaptivity and flexibility is entailed by the dynamic nature of wireless channels. Classical model-based receiver processing is inherently adaptive and flexible: The receiver periodically estimates the channel  using the available pilots, and then uses this estimate to adapt the operation of the receiver baseband chain, which is a direct function of the channel coefficients. In contrast,  for deep receivers, the dependence of the weights of the \ac{dnn} on the channel state is indirect, and hence designing flexible, channel-adaptive, \acp{dnn}-based processing  is a non-trivial task.  

Current state of the art on deep receivers encompasses the following three main approaches to address channel variations.
\begin{enumerate}[label={\em A\arabic*}]
\item \label{itm:joint} {\em Joint Learning}: The most straightforward approach amounts to optimizing a \emph{single} \ac{dnn} model to maximize performance \emph{on average} over a broad range of channel conditions.  Methods in this class train a \ac{dnn} using data corresponding to an extensive set of expected channel realizations, aiming to learn a mapping that is tailored to the distribution of the channel \cite{oshea2017introduction}.
Accordingly, joint learning may be thought of  as seeking  the optimal \emph{non-coherent receiver}, which is agnostic to the current channel realization. As a result, performance degradation as compared to a \emph{coherent receiver} is generally to be expected.


\item \label{itm:input} {\em Channel as Input}: An alternative approach  uses an instantaneous estimate of the channel as an additional input to the  \ac{dnn}~\cite{honkala2021deeprx}. Among the main drawbacks of this approach are the limited flexibility in accommodating different system dimensions, e.g., number of antennas or number of users, and the lack of structure in the way different inputs, such as received signals and channel state information, are handled.
\item \label{itm:online} {\em Online Training}: As  illustrated in Fig.~\ref{fig:online_training}, in online training, decoded data from prior blocks is used, alongside new pilots, to adapt the deep receiver to channel variations. This class of approaches inherits the limitations of \emph{continual learning}, such as catastrophic forgetting, and is generally not suitable to ensure fast adaptation.

\end{enumerate}

The mentioned shortcomings of the three existing approaches reviewed above motivate a fundamental rethinking of the application of machine learning tools to wireless receivers along the three directions illustrated in Fig.~\ref{tab:methods}.

\begin{itemize}
    \item The {\bf architecture} of the \ac{dnn} should be carefully selected on the basis of domain knowledge so as to reduce data requirements, while also ensuring efficient implementation of the model. This amounts to improvements in terms of the {\em inductive bias} on which learning is based.
    \item The {\bf data} used for learning should be augmented, when possible, by leveraging the inherent redundancies of encoded signals. 
    \item The {\bf training algorithm} should make use of historical data while also preparing for quick adaptation to changing channel conditions. 
\end{itemize}
In the following sections we review candidate approaches for each of these aspects, as summarized in Fig.~\ref{tab:methods}. 
	\begin{figure*}
    \centering
    \includegraphics[width=\linewidth]{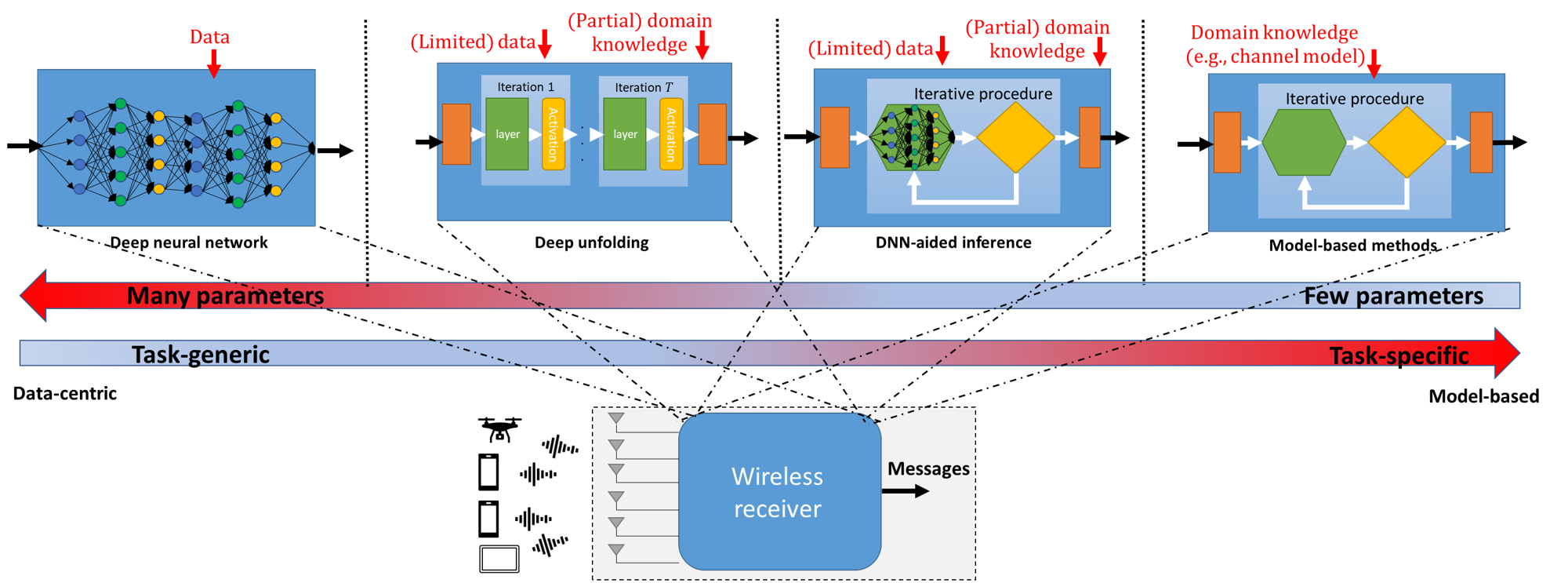}
    \caption{Illustration of model-based, data-driven, and model-based deep learning framework for deep receivers.}
    \label{fig:model_based_deep_learning}
\end{figure*}

\section{Architecture}
\label{sec:architecture}
The standard neural architectures employed in  \ac{ai} systems for communication are  based on highly-parameterized, unstructured, deep neural models such as feed-forward neural networks. The over-parameterization has been found to be advantageous in a host of other tasks, such as \ac{nlp}. However, since  deep receivers should adapt to time-varying conditions using limited training data, this type of architectures is typically undesirable. In this section, we introduce ways to design tailored model architectures by leveraging domain knowledge with the goal of improving adaptivity and data efficiency. In Sec.~\ref{sec:training}, we will also study data-driven approaches for the optimization of the inductive bias -- also known as meta-learning -- and see how they can be combined with model-driven architectures introduced in this section to further reduce the generalization gap.



In \emph{model-based deep learning}, \ac{dnn} architectures are designed  that are inspired by model-based algorithms tailored to the particular problem of interest~\cite{shlezinger2022model}. In the context of deep receivers, the dominant  model-based deep learning methodologies are  {\em deep unfolding} and {\em \ac{dnn}-aided inference}, which are illustrated in Fig.~\ref{fig:model_based_deep_learning} and discussed next. 

Many model-based algorithms used by wireless receivers rely on iterative optimizers that operate by gradually improving an optimization variable based on an objective function. {\bf Deep unfolding}  converts an iterative optimizer into a discriminative \ac{ai} model by introducing trainable parameters within each of a fixed number of iterations  \cite{shlezinger2022model}. Training a deep unfolding architecture can thus adapt an iterative optimizer on the basis of  available data for a given problem of interest. As we detail next, the aim is addressing model and/or algorithmic deficiencies of the original algorithm. 


Specifically, deep unfolding enhances iterative optimizers in the following ways (see~\cite{shlezinger2022model} for further details).
\begin{itemize}
    \item {\em Learned Hyperparameters}: Iterative optimizers often includes hyperparameters, such as step-sizes, damping factors, and regularization coefficients, that  are typically tuned by hand by the designer and shared among all iterations. Deep unfolding can treat such hyperparameters as trainable parameters. This is useful to cope with forms of {\em algorithm deficiency}, whereby an iterative algorithm requires too many iterations or struggles to converge to a suitable decision. For example, the work~\cite{he2018model} showed that  unfolding the orthogonal approximate message passing algorithm for \ac{mimo} detection, and learning iteration-dependent scaling coefficients, notably improves performance, requiring only a few iterations. 
    \item {\em Learned Objective}: Deep unfolding can also enhance an iterative algorithm by tuning the  objectives function approximately optimized at each iteration.  This optimization addresses  {\em algorithm deficiencies}, in a manner similar to the optimization of hyperparameters; as well as {\em model deficiencies} by adapting the design criterion to observed data, rather than to assumptions about the model.  A representative example is the MMNet architecture proposed in \cite{khani2020adaptive} for unfolding \ac{mimo} detection. MMNet, which is based on proximal gradient steps, parameterizes the gradient computation procedure at each iteration,  effectively using an iteration-dependent design objective.  
    \item {\em \ac{dnn} Conversion}: The third form of deep unfolding incorporates a full \ac{dnn} module within each iteration in order to implement some functionality of the solver in the most flexible manner.  \ac{dnn} conversion is  suitable for handling {\em model deficiency}, since the \ac{dnn} modules can learn how to best realize  model-independent internal computations at each iteration. For instance, DeepSIC proposed in \cite{shlezinger2019deepSIC} is derived from the iterative \ac{sic} \ac{mimo} detection algorithm with the introduction of \ac{dnn} models for implementing each stage of interference cancellation and soft detection in a manner agnostic to the underlying channel model. 
\end{itemize}

{\bf \ac{dnn}-Aided inference} refers to a family of model-based deep learning methods that incorporate \acp{dnn} into model-based methods that do not implement iterative processing. 
A representative example is the ViterbiNet  equalizer proposed in \cite{shlezinger2019viterbinet}. Viterbi equalization is applicable to any finite-memory channel, as long as one can compute the conditional distribution of channel output given the corresponding input, also known as likelihood. Based on this observation, ViterbiNet implements the Viterbi algorithm while using a \ac{dnn} to compute the likelihood. In this way, ViterbiNet addresses \emph{model deficiencies} by operating in a channel-model-agnostic manner and requiring only the conventional finite-memory modelling assumption to hold.


 
	\section{Data}
\label{sec:data}

The amount of data  obtained from pilots is typically insufficient to train an \ac{ai} model for a deep receiver. This motivates the introduction of strategies that \emph{expand} the available  labelled training data set without requiring the transmission of more pilots. As we detail in this section, existing techniques apply either self-supervised learning or data augmentation. 



\begin{figure}
    \centering
    \includegraphics[width=\columnwidth]{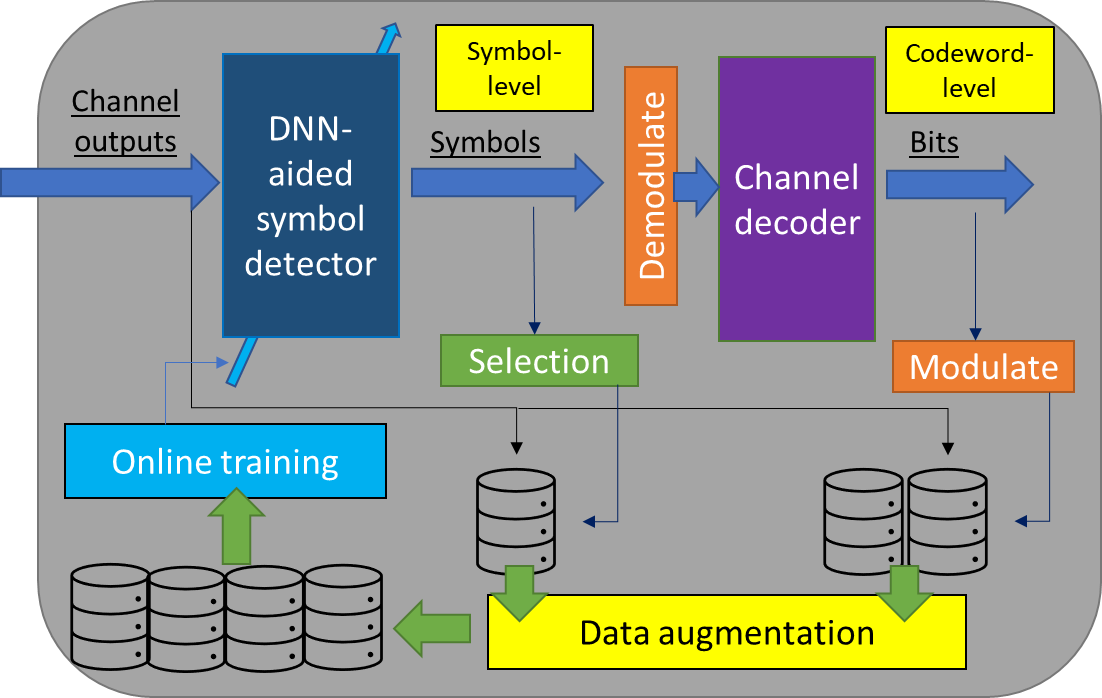}
    \caption{Data acquisition pipeline for deep receivers without harming spectral efficiency.}
    \label{fig:data_tasks}
\end{figure}

With  {\em self-supervised learning},  training data is extended using the redundancy of transmitted signals either at the symbol level or at the codeword level. In contrast, in  {\em data augmentation}, the goal is to enrich the given labelled data set by leveraging  invariance properties of the data. As summarized in Fig.~\ref{fig:data_tasks}, these approaches can be potentially combined, and integrated with a number of different  architectures (Sec.~\ref{sec:architecture}) and  training algorithms (Sec.~\ref{sec:training}). 




{\bf Codeword-level self-supervision} exploits the presence of channel coding to generate labelled data from channel outputs. It uses error correction codes to correct detection errors, and then utilizes the corrected data as labelled data for training, as long as  the codewords are decoded successfully \cite{shlezinger2019viterbinet, fischer2022adaptive,schibisch2018online}.  

{\bf Symbol-level self-supervision} obtains labelled data from information symbols without relying on channel decoding. This is useful since some symbols can be correctly detected even the decoding on the overall codeword fails. 
Symbol-level self-supervision hence requires reliable \emph{soft detection} measures to indicate the degree to which each information symbol may be considered to be correctly received~\cite{finish2022symbol}.  

{\bf Data augmentation} is an established framework in conventional \ac{ai} domains to enrich training sets by leveraging known invariances in the data. For instance, for image classification, one can use a single image to generate multiple images with the same label by rotating or clipping it.  While data augmentation is quite common in \ac{ai}, it is highly geared towards image and language data.  Data augmentations for digital communications have been  explored  in \cite{huang2019data}, and more recently in \cite{raviv2022data}. The techniques studied in \cite{raviv2022data}  include leveraging the symmetry in digital constellations to project error patterns between different symbols; exploiting the independence between the noise and the transmitted symbols to generate additional noisy realizations; and accounting for forms of invariance to constellation-preserving rotations exhibited by wireless channels.
 
	\section{Training}
\label{sec:training}

%

The training algorithm addresses the optimization of a data-dependent loss function, with the goal of identifying models with satisfactory generalization performance. The performance of a training algorithm depends, in practice, on (\emph{i}) the choice of the loss function; (\emph{ii}) the optimization algorithm; and (\emph{iii}) the relevance and quality of the data used to evaluate the training loss.  In this section, we review communication-oriented approaches for designing adaptive, data-efficient, training algorithm for deep receivers based on (\emph{i}) meta-learning, (\emph{ii}) generalized Bayesian learning, and (\emph{iii}) modular learning.

{\bf Meta-learning} is a general framework that seeks to obtain a \emph{data-efficient} training procedure that is applicable for multiple
tasks of interest \cite{chen2023learning}. A training procedure that is data-, or sample-, efficient is able to achieve  a small generalization gap, while using a small amount of training data. Meta-learning and model-based learning (see Sec.~\ref{sec:architecture}) are two complementary approaches that reduce the generalization gap under a fixed amount of training data: The former is data-driven and typically optimizes the training algorithm;  while the latter is model-driven and optimizes the architecture. While meta-learning encompasses a variety of conceptually distinct methods,  the  prominent approaches for application to deep receivers are gradient-based meta-learning and hypernetwork-based meta-learning. 
\begin{itemize}
    \item {\em Gradient-based meta-learning}: Gradient-based meta-learning  optimizes some of the hyperparameters of a first-order training algorithm. While in principle, one could ``meta-learn'' any hyperparameter, such as the learning rate, optimizing the \emph{initial weights} of the \acp{dnn} has been found  to be extremely beneficial for boosting  adaptation and flexibility of training procedures in many applications, including in wireless communications~\cite{chen2023learning}. \ac{dnn} initialization is a form of inductive bias, since the parametric function space of the \ac{dnn} becomes restricted by enforcing adherence to the initialization through a limited number of gradient-based updates. Meta-learning can be combined with a model-based inductive bias, as demonstrated in  \cite{raviv2023online}.
    \item {\em Hypernetwork-based meta-learning}: Gradient-based meta-learning requires running a number of (stochastic) gradient updates. An alternative approach that does not require in real-time any additional  optimization for adaptation to new tasks incorporates a \emph{hypernetwork} in the system, alongside the main \ac{dnn}. The hypernetwork takes as input the available data, or any other context information,  regarding the task of interest, and produces at the output the weights of the main \ac{dnn}. More precisely, typically, only a subset of weights of the main \ac{dnn} are updated; and/or each output of the hypernetwork  affects simultaneously a group of weights, e.g., in the same layer, of the main \ac{dnn}.  Hypernetwork-based meta-learning has been applied successfully in wireless communication systems, including for beamforming and \ac{mimo} detection \cite{goutay2020deep,liu2022learning}.
\end{itemize}

{\bf Bayesian learning} is the gold standard for training strategies that  aim at producing  AI models offering  a \emph{reliable} assessment of the  uncertainty of their decisions.  Such reliable AI models must output  confidence measures that reflect the true accuracy of their  decisions. Bayesian learning boosts  reliability by treating the model parameters as random variables, and by accordingly  maintaining  a \emph{distribution} over the weights of a \ac{dnn}. This distribution is meant to capture \emph{epistemic uncertainty} in the presence of limited training data. 

Bayesian learning involves particle-based, deterministic or stochastic, procedures, or optimization over the parameters of the distribution in the model parameter space. Such optimization addresses a training criterion that includes an information-theoretic regularizer enforcing closeness to a prior distribution. 


For deep receivers, boosting the reliability of a \ac{dnn} model allows the latter to provide informative soft decision to downstream \ac{dnn} or model-based modules, e.g., for soft decoding. This makes it possible for the different modules of a deep receiver to ``trust'' the outputs of other modules~\cite{raviv2023modular}.

Generalized forms of Bayesian learning allow for a flexible choice of  the regularization function, as well as of the  data-fitting part of the training objective. Such methods were shown to be useful in wireless systems for their capacity to deal with model misspecification and outliers \cite{zecchin2022robust}.  

{\bf Modular learning} exploits the interpretable structure of hybrid model-based deep receivers to facilitate rapid learning from limited data. As opposed to meta-learning and Bayesian learning, modular learning is specific to model-based deep learning architectures. It builds on the fact that, unlike black-box \acp{dnn}, in model-based deep learning architectures, one can often assign a concrete functionality to different trainable sub-modules of the architecture, and not just to its input and output.  
Each functionality may then be adapted at different rates and times, as some functionalities may require rapid adaptation, while the others may be kept unchanged over a longer time scale.


This approach was applied in \cite{raviv2023online} for online adaptation of the DeepSIC \ac{mimo} receiver of \cite{shlezinger2019deepSIC}.  There, the ability to associate different users with sub-modules of the deep receivers  was leveraged to carry out the online training of  sub-modules associated with  users that are identified as being characterized by faster dynamics. The method was shown to dramatically reduce the number of gradient-based updates and the amount of data needed for online training.   

	\section{Numerical Results}
\label{sec:results}

In this section we showcase the impact of schemes designed to facilitate light-weight, adaptive, and flexible \ac{ai} across the three AI pillars highlighted throughout this article. We focus on finite-memory \ac{siso} channels (with $4$ taps) and memoryless $4\times 4$ multi-user \ac{mimo} time-varying channels with \ac{bpsk} and \ac{qpsk} symbols, respectively\footnote{The source code used in our experiments is available at \href{https://github.com/tomerraviv95/facilitating-adaptation-deep-receivers}{https://github.com/tomerraviv95/facilitating-adaptation-deep-receivers}}. 



\begin{figure*}[t]
    \centering
    \begin{subfigure}[b]{0.48\textwidth}
    \includegraphics[width=\textwidth,height=0.24\textheight]{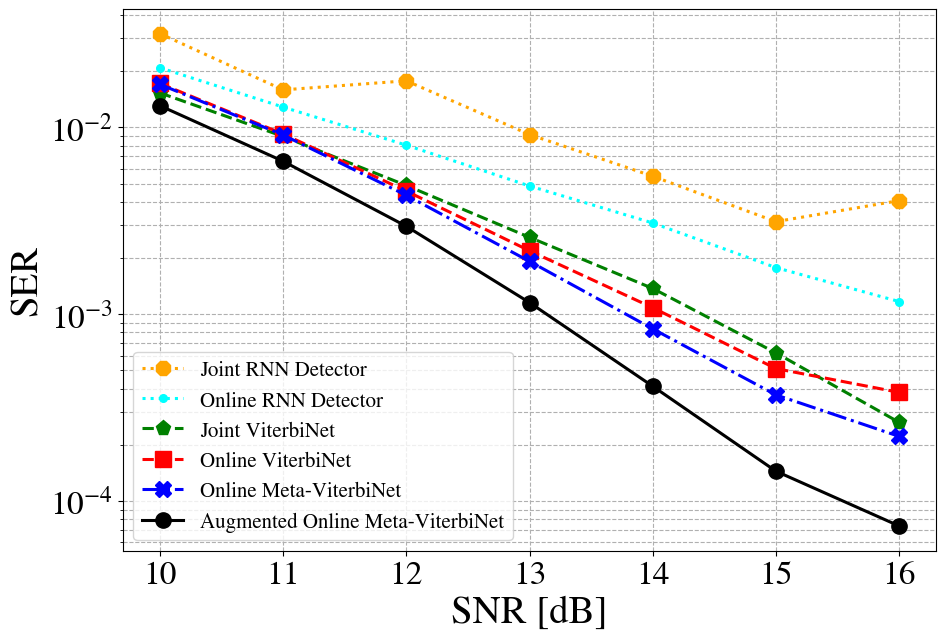}
    \caption{\ac{siso} - \acs{ser} as SNR.}
    \label{fig:siso_ber}
    \end{subfigure}
    \begin{subfigure}[b]{0.48\textwidth}
    \includegraphics[width=\textwidth,height=0.24\textheight]{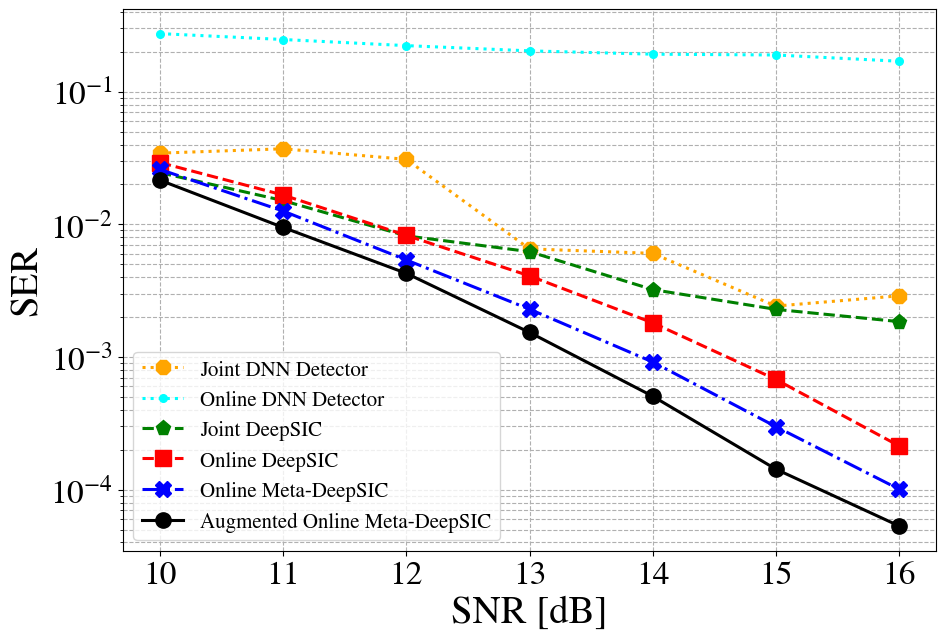}
    \caption{\ac{mimo} - \acs{ser} vs. SNR.}
    \label{fig:mimo_ber}
    \end{subfigure}
    \caption{Average \acs{ser} after transmission of $300$ blocks in a time-varying channel as a function of \acs{snr}.}
    \label{fig:compared_methods} 
    \figSpace
\end{figure*}

{\bf Architecture:} 
In each channel, we consider a model-based \ac{dnn} architecture, as well as black-box \ac{dnn}, having roughly three times more parameters. 
For the \ac{siso} channel, with a finite channel memory of $L$ symbols, we compare  ViterbiNet \cite{shlezinger2019viterbinet} with a \ac{rnn}-based  symbol detector with  a window size of $L$, followed by a linear layer and the softmax function. For the \ac{mimo} channel, the DeepSIC receiver \cite{shlezinger2019deepSIC} with three iterations is compared to a fully connected  \ac{dnn} composed of four layers with ReLU activations followed by the softmax layer.

{\bf Data:}  
For each coherence duration, $200$ pilot symbols are available. 
We compare standard training with training that leverages data augmentation. For the latter scheme,  
at each time step, the pilot data is enriched with $600$ artificial symbols via  a constellation-conserving projection and a translation-preserving transformation~\cite{raviv2022data}. 

{\bf Training:} 
We consider the following training methodologies:
\begin{itemize}
    \item {\em Joint training}: The receiver is trained offline, using $5000$ symbols simulated from a multitude of channel realizations. 
    No additional training is done at run time. 
    \item {\em Online training}: The receiver is trained initially using $200$ symbols, and then it adapts  online  by utilizing either the pilot data or the augmented pilot data.
    \item {\em Online meta-learning}: The training algorithm is optimized via meta-learning that utilizes accumulated training data from previous channel realizations, while adaptation takes place via few gradient-based updates from the online meta-learned initialization~\cite{raviv2023online}. 
\end{itemize}





{\bf Results:}
Fig.~\ref{fig:siso_ber} and Fig.~\ref{fig:mimo_ber}  depict the average  \ac{ser} as a function of \ac{snr}. While standard black-box models suffer from large generalization gaps due to the limited availability of training data, deep receivers with model-based {architectures}, namely ViterbiNet  \cite{shlezinger2019viterbinet} and DeepSIC  \cite{shlezinger2019deepSIC}, demonstrate successful detection performance by adapting to the time-varying channel in an online manner. 

The performance is seen to be further improved by optimizing the \emph{training algorithm} via meta-learning, as well as by increasing the \emph{data} size via data augmentation. Overall,  these results indicate that the reviewed methods are complementary, contributing to the challenges of adapting to time-varying channels in different ways. This leads to the conclusion that  designing AI models for communications can benefit from a rethinking of deep learning tools across all three \ac{ai} pillars.



	\section{Future Research Directions}
\vspace{-0.1cm}
\label{sec:conclusion}

We conclude by identifying some representative  directions for future research.  


\subsection{Deciding When to Train}
\label{subsec:drift_detection}

The schemes surveyed thus far are geared towards enabling efficient online on-device training. In this regard, a key open question 
 is how to determine {\em when} to train online. Periodically re-training, e.g., at  each coherence period, may be excessively complex, particularly when channel variations are relatively smooth. Efficient deep receiver would benefit from monitoring mechanisms that can determine when to adapt the model and/or when to meta-learn the inductive bias. One possible way to achieve this goal is via data drift detection, a topic widely studied in the machine learning literature~\cite{lu2018learning}. 
 While some basic drift detection mechanisms can be directly applied to communication systems, advanced mechanisms that leverage communication-specific characteristics may require further development.

\subsection{Fitting the Architecture to the Scenario}
\label{subsec:adaptive_pruning_quantization}

Deep receivers are often composed of multiple layers, wherein each element takes part in the computation. Thus, even for relatively light-weight architectures, full model computations may incur computational overhead exceeding the limited resources available, particularly for some edge devices. 
This problem is typically tackled via pruning methods, which aim to bridge the complexity-performance gap by removing redundant parts of the models. While most existing pruning methods find a single, input-independent, light-weight model, for wireless communication systems it may be preferable to adopt \emph{input-dependent, adaptive} pruning methods~\cite{singh2019play}, that can adapt complexity to the current requirements. 

\subsection{ Hardware-Aware AI}
The schemes surveyed in this paper do not make use of any special characteristics of the hardware available at the host device, focusing instead of generic improvements based on limiting the architecture parameterization and/or the number of training iterations. Larger efficiency gains are to be expected with \ac{ai} methods that are aware of the specific hardware at the wireless receiver, which may encompass different  technologies such as emerging in-memory computing chips. 


\subsection{Continual Bayesian Learning}
Bayesian learning was introduced in this article as a promising solution for  deep receivers thanks to the potential gains that are enabled by the deployment of more reliable AI modules. Another important advantage of Bayesian learning is its capacity to support continual learning via the update of the model parameter distribution~\cite{chang2022diagonal}.
The integration of online adaptation with Bayesian learning may further enhance the performance of deep receivers.

\subsection{Collaborative Learning and Inference}
\label{subsec:collaboration}
As mentioned, deep receivers are practically constrained by the hardware available at the host device. Since deep receivers are likely to be deployed in environments containing other, similar, devices, this limitation may be mitigated via resource sharing across devices. Such collaboration may entail the exchange of data and/or model information, and it may be supported by device-to-device communication capabilities. This idea is deeply connected to federated learning ~\cite{gafni2022federated} and collaborative inference~\cite{shlezinger2022collaborative}.

	\bibliographystyle{IEEEtran}
	\bibliography{IEEEabrv,refs}

\begin{thebibliography}{10}
\providecommand{\url}[1]{#1}
\csname url@samestyle\endcsname
\providecommand{\newblock}{\relax}
\providecommand{\bibinfo}[2]{#2}
\providecommand{\BIBentrySTDinterwordspacing}{\spaceskip=0pt\relax}
\providecommand{\BIBentryALTinterwordstretchfactor}{4}
\providecommand{\BIBentryALTinterwordspacing}{\spaceskip=\fontdimen2\font plus
\BIBentryALTinterwordstretchfactor\fontdimen3\font minus
  \fontdimen4\font\relax}
\providecommand{\BIBforeignlanguage}[2]{{%
\expandafter\ifx\csname l@#1\endcsname\relax
\typeout{** WARNING: IEEEtran.bst: No hyphenation pattern has been}%
\typeout{** loaded for the language `#1'. Using the pattern for}%
\typeout{** the default language instead.}%
\else
\language=\csname l@#1\endcsname
\fi
#2}}
\providecommand{\BIBdecl}{\relax}
\BIBdecl

\bibitem{samsung202065}
``{6G} - the next hyper connected experience for all,'' \emph{Samsung 6G
  Vision}, 2020.

\bibitem{dai2020deep}
L.~Dai, R.~Jiao, F.~Adachi, H.~V. Poor, and L.~Hanzo, ``Deep learning for
  wireless communications: An emerging interdisciplinary paradigm,''
  \emph{{IEEE} Wireless Commun.}, vol.~27, no.~4, pp. 133--139, 2020.

\bibitem{tong2022nine}
W.~Tong and G.~Y. Li, ``Nine challenges in artificial intelligence and wireless
  communications for {6G},'' \emph{{IEEE} Wireless Commun.}, vol.~29, no.~4,
  pp. 140--145, 2022.

\bibitem{he2018model}
H.~He, C.-K. Wen, S.~Jin, and G.~Y. Li, ``Model-driven deep learning for {MIMO}
  detection,'' \emph{{IEEE} Trans. Signal Process.}, vol.~68, pp. 1702--1715,
  2020.

\bibitem{khani2020adaptive}
M.~Khani, M.~Alizadeh, J.~Hoydis, and P.~Fleming, ``Adaptive neural signal
  detection for massive {MIMO},'' \emph{{IEEE} Trans. Wireless Commun.},
  vol.~19, no.~8, pp. 5635--5648, 2020.

\bibitem{shlezinger2019deepSIC}
N.~Shlezinger, R.~Fu, and Y.~C. Eldar, ``{DeepSIC}: Deep soft interference
  cancellation for multiuser {MIMO} detection,'' \emph{{IEEE} Trans. Wireless
  Commun.}, vol.~20, no.~2, pp. 1349--1362, 2021.

\bibitem{shlezinger2019viterbinet}
N.~Shlezinger, N.~Farsad, Y.~C. Eldar, and A.~J. Goldsmith, ``{ViterbiNet}: A
  deep learning based {Viterbi} algorithm for symbol detection,'' \emph{{IEEE}
  Trans. Wireless Commun.}, vol.~19, no.~5, pp. 3319--3331, 2020.

\bibitem{fischer2022adaptive}
M.~B. Fischer, S.~D{\"o}rner, S.~Cammerer, T.~Shimizu, H.~Lu, and S.~Ten~Brink,
  ``Adaptive neural network-based {OFDM} receivers,'' in \emph{Proc. IEEE
  SPAWC}, 2022.

\bibitem{schibisch2018online}
S.~Schibisch, S.~Cammerer, S.~D{\"o}rner, J.~Hoydis, and S.~ten Brink, ``Online
  label recovery for deep learning-based communication through error correcting
  codes,'' in \emph{Proc. IEEE ISWCS}, 2018.

\bibitem{finish2022symbol}
R.~A. Finish, Y.~Cohen, T.~Raviv, and N.~Shlezinger, ``Symbol-level online
  channel tracking for deep receivers,'' in \emph{Proc. IEEE ICASSP}, 2022, pp.
  8897--8901.

\bibitem{huang2019data}
L.~Huang, W.~Pan, Y.~Zhang, L.~Qian, N.~Gao, and Y.~Wu, ``Data augmentation for
  deep learning-based radio modulation classification,'' \emph{{IEEE} Access},
  vol.~8, pp. 1498--1506, 2019.

\bibitem{raviv2022data}
T.~Raviv and N.~Shlezinger, ``Data augmentation for deep receivers,''
  \emph{{IEEE} Trans. Wireless Commun.}, 2023, early access.

\bibitem{raviv2023online}
T.~Raviv, S.~Park, O.~Simeone, Y.~C. Eldar, and N.~Shlezinger, ``Online
  meta-learning for hybrid model-based deep receivers,'' \emph{{IEEE} Trans.
  Wireless Commun.}, 2023, early access.

\bibitem{goutay2020deep}
M.~Goutay, F.~A. Aoudia, and J.~Hoydis, ``Deep hypernetwork-based {MIMO}
  detection,'' in \emph{Proc. IEEE SPAWC}, 2020.

\bibitem{chen2023learning}
L.~Chen, S.~T. Jose, I.~Nikoloska, S.~Park, T.~Chen, and O.~Simeone, ``Learning
  with limited samples: Meta-learning and applications to communication
  systems,'' \emph{Foundations and {T}rends{\textregistered} in {S}ignal
  {P}rocessing}, vol.~17, no.~2, pp. 79--208, 2023.

\bibitem{liu2022learning}
Y.~Liu and O.~Simeone, ``Learning how to transfer from uplink to downlink via
  hyper-recurrent neural network for {FDD} massive {MIMO},'' \emph{{IEEE}
  Trans. Wireless Commun.}, vol.~21, no.~10, pp. 7975--7989, 2022.

\bibitem{zecchin2022robust}
M.~Zecchin, S.~Park, O.~Simeone, M.~Kountouris, and D.~Gesbert, ``Robust
  {B}ayesian learning for reliable wireless {AI}: Framework and applications,''
  \emph{{IEEE} Trans. on Cogn. Commun. Netw.}, 2023, early access.

\bibitem{raviv2023modular}
T.~Raviv, S.~Park, O.~Simeone, and N.~Shlezinger, ``Modular model-based
  {B}ayesian learning for uncertainty-aware and reliable deep {MIMO}
  receivers,'' in \emph{Proc. IEEE ICC}, 2023.

\bibitem{shlezinger2022model}
N.~Shlezinger, Y.~C. Eldar, and S.~P. Boyd, ``Model-based deep learning: On the
  intersection of deep learning and optimization,'' \emph{{IEEE} Access},
  vol.~10, pp. 115\,384--115\,398, 2022.

\bibitem{oshea2017introduction}
T.~O’Shea and J.~Hoydis, ``An introduction to deep learning for the physical
  layer,'' \emph{{IEEE} Trans. on Cogn. Commun. Netw.}, vol.~3, no.~4, pp.
  563--575, 2017.

\bibitem{honkala2021deeprx}
M.~Honkala, D.~Korpi, and J.~M. Huttunen, ``Deep{R}x: Fully convolutional deep
  learning receiver,'' \emph{{IEEE} Trans. Wireless Commun.}, vol.~20, no.~6,
  pp. 3925--3940, 2021.

\bibitem{lu2018learning}
J.~Lu, A.~Liu, F.~Dong, F.~Gu, J.~Gama, and G.~Zhang, ``Learning under concept
  drift: A review,'' \emph{IEEE Transactions on Knowledge and Data
  Engineering}, vol.~31, no.~12, pp. 2346--2363, 2018.

\bibitem{singh2019play}
P.~Singh, V.~K. Verma, P.~Rai, and V.~P. Namboodiri, ``Play and prune: Adaptive
  filter pruning for deep model compression,'' \emph{arXiv preprint
  arXiv:1905.04446}, 2019.

\bibitem{chang2022diagonal}
P.~G. Chang, K.~P. Murphy, and M.~Jones, ``On diagonal approximations to the
  extended {K}alman filter for online training of {B}ayesian neural networks,''
  in \emph{Continual Lifelong Learning Workshop at ACML 2022}, 2022.

\bibitem{gafni2022federated}
T.~Gafni, N.~Shlezinger, K.~Cohen, Y.~C. Eldar, and H.~V. Poor, ``Federated
  learning: A signal processing perspective,'' \emph{{IEEE} Signal Process.
  Mag.}, vol.~39, no.~3, pp. 14--41, 2022.

\bibitem{shlezinger2022collaborative}
N.~Shlezinger and I.~V. Baji{\'c}, ``Collaborative inference for {AI}-empowered
  {IoT} devices,'' \emph{IEEE Internet of Things Magazine}, vol.~5, no.~4, pp.
  92--98, 2022.

\end{thebibliography}

 \begin{IEEEbiographynophoto}{Tomer Raviv} (tomerraviv95@gmail.com)
is currently pursuing his Ph.D degree in electrical engineering 
in Ben-Gurion University.
\end{IEEEbiographynophoto}	
\vskip -2\baselineskip plus -1fill

 \begin{IEEEbiographynophoto}{Sangwoo Park} (sangwoo.park@kcl.ac.uk)
is currently a research associate at the Department of Engineering, King's Communications, Learning and Information Processing (KCLIP) lab, King's College London, United Kingdom.
\end{IEEEbiographynophoto}	
\vskip -2\baselineskip plus -1fill

 \begin{IEEEbiographynophoto}{Osvaldo Simeone} (osvaldo.simeone@kcl.ac.uk)
is a Professor of Information Engineering with the Centre for Telecommunications Research at the Department of Engineering of King's College London, where he directs the King's Communications, Learning and Information Processing lab.
\end{IEEEbiographynophoto}	
\vskip -2\baselineskip plus -1fill

 \begin{IEEEbiographynophoto}{Yonina C. Eldar} (yonina.eldar@weizmann.ac.il)
is a Professor in the Department of Math and Computer Science, Weizmann Institute of Science, Israel, where she heads the center for Biomedical Engineering and Signal Processing. She is a member of the Israel Academy of Sciences and Humanities, an IEEE Fellow and a EURASIP Fellow.
\end{IEEEbiographynophoto}	
\vskip -2\baselineskip plus -1fill

\begin{IEEEbiographynophoto}{Nir Shlezinger} (nirshl@bgu.ac.il) is an Assistant Professor in the School of Electrical and Computer Engineering in Ben-Gurion University, Israel. 
\end{IEEEbiographynophoto}
\vskip -2\baselineskip plus -1fil

\end{document}